\documentclass[a4paper,11pt]{article}
\usepackage{pos}
\usepackage{comment}
\newcommand{\bs}[1]{\boldsymbol{#1}}

\newcommand{\pd}[2]{\frac{\partial #1}{\partial #2}}

\title{Taming NSPT fluctuations in $O(N)$ Non-Linear Sigma Model: simulations in the large $N$ regime}
\ShortTitle{Taming NSPT fluctuations in $O(N)$ Non-Linear Sigma Model}

\author*[a,b]{Paolo Baglioni}
\author[a,b]{Francesco Di Renzo}

\affiliation[a]{University of Parma,\\
Parco Area delle Scienze, 7/A, Parma, Italy}

\affiliation[b]{INFN gruppo collegato di Parma,\\
Parco Area delle Scienze, 7/A, Parma, Italy}

\emailAdd{paolo.baglioni@unipr.it}
\emailAdd{francesco.direnzo@unipr.it}

\abstract{The Non-Linear Sigma Model (NLSM) is an example of a field theory on a target space exhibiting intricate geometry. One remarkable characteristic of the NLSM is asymptotic freedom, which triggers interest in perturbative calculations. In the lattice formulation of NLSM, one would naturally rely on Numerical Stochastic Perturbation Theory (NSPT) to conduct high-order computations. However, when dealing with low-dimensional systems, NSPT reveals increasing statistical fluctuations with higher and higher orders. This of course does not come as a surprise and one is ready to live with this, as long as the noise is not going to completely kill the signal, which unfortunately in some models does take place. We investigate how, in the $O(N)$ context, this behaviour strongly depends on $N$. As expected, larger $N$ values make higher-order computations feasible.}

\FullConference{European network for Particle physics, Lattice field theory and Extreme computing (EuroPLEx2023)\\
 11-15 September 2023\\
Berlin, Germany\\}

\begin{document}
\maketitle

\section{Introduction and motivations}

Lattice Gauge Theories (LGT) are an example of regularized field theories designed for a non-perturbative approach. Through Monte Carlo simulations, LGTs remain one of the primary tools for numerically investigating non-perturbative physics. In this context, perturbation theory is known to be particularly cumbersome. Nevertheless, perturbation theory holds its significance even in lattice theories. Numerical Stochastic Perturbation Theory (NSPT) \cite{DiRenzo1994} stands out as a renowned computational tool for generating perturbative expansions in lattice theories, sharing striking similarities with non-perturbative Monte Carlo methods \cite{DiRenzo2004}.

NSPT is a fully automated perturbative Monte Carlo algorithm that involves order-by-order numerical integration of stochastic differential equations. It has gained considerable interest among lattice practitioners mainly because of the agility of its implementation compared to its significant potential. Since it has been introduced, NSPT simulations have allowed the calculation of high-order loop corrections in lattice QCD \cite{DiRenzo2007, DiRenzo2008, DiRenzo2010, DiRenzo2011, DallaBrida20172} and also across various other theories \cite{Kitano2021, GonzlezArroyo2019, GonzlezArroyo2022}. Currently, NSPT has been implemented in different forms \cite{DallaBrida2017, DallaBrida20172} and remains one of the main distinctive tools for investigating asymptotic perturbative behavior in lattice QCD, probing fundamental properties such as infrared renormalons \cite{Bauer2012, Bali2013, Bali2014, DelDebbio2018}. 

One interesting aspect of NSPT is the freedom to expand around different non-trivial vacua. In this respect, low-dimensional models emerge as the optimal candidates for exploratory analysis, but apart from first-order loop computations in quantum mechanics \cite{Baglioni2023}, perturbative expansions on top of instanton-like vacua still remain mostly unexplored. As a matter of fact, it is well-known that significant fluctuations arise in low-dimensional models even at relatively low perturbative orders, precluding straightforward NSPT implementation for such computations \cite{Alfieri2000}. A reasonable explanation is that fluctuations emerge due to the limited number of degrees of freedom. This hypothesis is supported by the absence of such fluctuations in more complex systems like lattice QCD.

In this study, we consider a class of highly interesting low-dimensional lattice models, the $O(N)$ Non-Linear Sigma Model (NLSM) in $2D$. Besides its theoretical and phenomenological significance, the $O(N)$ NLSM provides an ideal testing ground for the aforementioned conjecture. In fact, we can change the number of local degrees of freedom by tuning the value of $N$. On very general grounds, we expect that fluctuations are less severe in the large $N$ regime, where the theory includes a large number of degrees of freedom. In this manuscript we report the current status of our simulations, which strongly suggests that the signals become increasingly free of fluctuations as the value of $N$ grows (see also \cite{baglioni2024, Baglioni2024c} for more details).  

\section{Numerical Stochastic Perturbation Theory}

NSPT has been inspired by the famous work of Parisi and Wu on Stochastic Quantization (SQ) \cite{Parisi1980}. The main idea is to translate the stochastic perturbation theory in a set of order-by-order differential equations that must be solved numerically, formulating a strategy for a perturbative Monte Carlo algorithm. 

In the SQ framework, we consider the degrees of freedom as function of an extra parameter, the \textit{stochastic time}:
\begin{align}
    \phi(\bs{x}) \rightarrow \phi(\bs{x},\tau)\, .
\end{align}
The evolution takes place in the stochastic time following the Langevin equation
\begin{align}
	\label{eq:langevin}
	\frac{d \phi(\bs{x},\tau)}{d\tau} = -\frac{\partial\mathcal{S}_E[\phi]}{\partial\phi(\bs{x},\tau)} + \eta(\bs{x},\tau) \, ,
\end{align}
where $\mathcal{S}_E[\phi]$ is the Euclidean action, and $\eta(\bs{x},\tau)$ is a white noise normalized such that
\begin{align}
	\label{eq:whiteNoise}
	\langle \eta(\bs{x},\tau) \rangle_\eta = 0 \, ,\quad  \langle \eta(\bs{x},\tau)\eta(\bs{x}',\tau') \rangle_\eta = 2\delta(\bs{x}-\bs{x}')\delta(\tau-\tau') \, .
\end{align}
In Eq.~\eqref{eq:whiteNoise} the subscript $\langle \ldots \rangle_{\eta}$ means the average over all possible white noise realizations, namely
\begin{align}
    \label{eq:averageOverNoise}
    \langle \dots \rangle_\eta = \frac{\int \mathcal{D}\eta \ \dots \ e^{-\frac{1}{4} \int d\bs{x} d\tau \eta^2(\bs{x},\tau)} }{\int \mathcal{D}\eta \ e^{-\frac{1}{4} \int d\bs{x} d\tau \eta^2(\bs{x},\tau)}} \, .
\end{align}
The central assertion of SQ is that the averaging over noise realizations at the limit of infinite stochastic time results in the expectation value of the corresponding Euclidean field theory, namely
\begin{align}
	\label{eq:convergence}
	\lim_{\tau\to\infty} \langle O[\phi(\bs{x}_1,\tau) , \phi(\bs{x}_2,\tau), \dots, \phi(\bs{x}_n,\tau)]\rangle_\eta = \langle O[\phi(\bs{x}_1), \phi(\bs{x}_2), \dots, \phi(\bs{x}_n)]\rangle\, ,
\end{align}
where
\begin{align}
	\langle O[\phi(\bs{x}_1), \phi(\bs{x}_2), \dots, \phi(\bs{x}_n)]\rangle\ = \frac{\int \mathcal{D}\phi\ O[\phi(\bs{x}_1), \phi(\bs{x}_2), \dots, \phi(\bs{x}_n)]\ e^{-\mathcal{S}_E[\phi]}}{\int \mathcal{D}\phi\ e^{-\mathcal{S}_E[\phi]} } \, .
\end{align}
NSPT involves formal expansion in the field's coupling constant, namely
\begin{align}
\label{eq:ptExpField}
    \phi(\bs{x},\tau) = \sum_{n=0}^{\infty}g^n \phi^{(n)}(\bs{x},\tau) \, ,
\end{align}
and the order-by-order numerical integration of the set of equations that are obtained by inserting the previous definition in Eq.~\eqref{eq:langevin}, i.e.
\begin{align}
\label{eq:orderByOrderLangevin}
\begin{split}
     \frac{d \phi^{(0)}(\bs{x},\tau)}{d\tau} & = -G_0^{-1}\phi^{(0)}(\bs{x},\tau) + \eta(\bs{x},\tau) \, , \\
     \frac{d \phi^{(1)}(\bs{x},\tau)}{d\tau} & = -G_0^{-1}\phi^{(1)}(\bs{x},\tau) + D_1(\phi^{(0)}) \, , \\
     \dots\\
      \frac{d \phi^{(n)}(\bs{x},\tau)}{d\tau} & = -G_0^{-1}\phi^{(n)}(\bs{x},\tau) + D_n(\phi^{(0)},\phi^{(1)}, \dots, \phi^{(n-1)}) \, .
\end{split}
\end{align}
In the above equation $G_0$ represents the free propagator, while $D_n$ denotes functions of the perturbative fields that couple different perturbative orders. It is important to emphasize that the NSPT code only needs to handle the Langevin dynamics in Eq.~\eqref{eq:langevin} {\em as it is}: all the order-by-order operations are automatically decoded using polynomial operations in the coupling constant (a feature that has recently gained popularity as Automatic Differentiation \cite{Catumba2023}). In addition, the observables are implemented in their non-perturbative form, with the corresponding perturbative expansion being automatically generated as
\begin{align}
\begin{split}
    \label{eq:nsptObservable}
    O[\phi] & = O^{(0)}[\phi^{(0)}] + g O^{(1)}[\phi^{(0)}, \phi^{(1)}] + g^2 O^{(2)}[\phi^{(0)}, \phi^{(1)},\phi^{(2)}] \, ,\\
    & = \sum_{n=0}^{\infty}g^n O^{(n)}[\phi^{(0)},\ldots,\phi^{(n)}] \, .
\end{split}
\end{align}
The numerical integration of Eqs.~\eqref{eq:orderByOrderLangevin} requires a designated small time step $\Delta\tau = \tau / n_{\text{steps}}$. The resulting discrete dynamics no longer satisfy Eq.~\eqref{eq:convergence}, requiring the continuum stochastic time extrapolation $\Delta\tau \to 0$ (a detailed presentation concerning the continuum stochastic time extrapolation is reported in \cite{Baglioni2024c}).

\section{\texorpdfstring{O(N)}{} NLSM: Lattice perturbation theory setup}

In this work, we study the $2D$ $O(N)$ Non-Linear Sigma Model defined by the action
\begin{align}
	S = -\frac{1}{g}\sum_{x,\mu} \bs{s}_x \cdot \bs{s}_{x+\mu} \, ,
\end{align}
where $\bs{s}_x$ is a $N$-component real scalar field constrained by $\bs{s}_x \cdot \bs{s}_x = 1$, and $g$ is the coupling constant. NLSMs are a very popular laboratory for studying particle and field properties in various contexts (we refer to \cite{zinn-justin2002} for a complete introduction to this topic). Here we are not interested in such features, but we will study statistical proprieties of the NSPT distributions as a function of $N$, which represents the number of local degrees of freedom. 

Perturbation theory is formulated in terms of the unconstrained field $\bs{\pi}_x$ \cite{elitzur1983}, defined by
\begin{align}
	\label{eq:stopi}
	\bs{s}_x = (\sqrt{g}\bs{\pi}_x,\sigma_x)\, ,
\end{align}
where the last component $\sigma_x$ is integrated out using the local constraint. Neglecting irrelevant contributions in perturbation theory \cite{elitzur1983}, the partition function is then expressed as
\begin{align}
	\label{eq:PF_PT}
	Z = \lim_{\lambda\to 0}\int \prod_{x}\ d\bs{\pi}_x \ e^{-\frac{1}{2}\sum_{x,\mu} \Bigl[ (\Delta_\mu\bs{\pi}_x)^2 + \lambda^2 \bs{\pi}_x^2 -\frac{1}{g}(\Delta_\mu\sqrt{1-g\bs{\pi}_x^2})^2 \Bigr] - \frac{1}{2}\sum_x \log{(1-g\bs{\pi}_x^2)}}\, .
\end{align}
In Eq.~\eqref{eq:PF_PT}, we introduced the infrared regulator $\lambda$, while $\Delta_\mu$ represents the usual lattice-discretized derivative. It is important to note that the action now involves logarithmic and square root interactions that need to be managed in the framework of perturbation theory. In this case, we have computed the Taylor expansions in the arguments and reorganized the resulting series in formal power series of the coupling constant.

A well-known observable in perturbation theory is the nearest-neighbor two-point function, namely the energy of the system:
\begin{align}
\begin{split}
	\label{eq:energy}
    E & = -\frac{1}{2V} \pd{\log{Z}}{\Bigl(\frac{1}{g}\Bigr)} = \frac{1}{2V} \sum_{x,\mu}\langle \bs{s}_x \cdot \bs{s}_{x+\mu} \rangle  \, , \\
    & = \frac{1}{2V} \sum_{x,\mu} \Bigl[ g \langle \bs{\pi}_x \cdot \bs{\pi}_{x+\mu} \rangle + \langle \sqrt{1 - g\bs{\pi}^2_x}\sqrt{1 - g\bs{\pi}^2_{x+\mu}} \rangle \Bigr]  \, , \\
    & = E^{(0)} + g E^{(1)} + g^2 E^{(2)} + ... + g^n E^{(n)} + ... \, .
\end{split}
\end{align}
In the above equation, the square roots require Taylor series expansion in the coupling constant as before. It turns out that the diagrammatic lattice perturbation theory is impractical due to the large number of new vertices generated at each perturbative order and only the first four loop corrections are known \cite{elitzur1983, alles1997}. In the following, we will compute perturbative corrections in Eq.~\eqref{eq:energy} using NSPT simulations.

\section{Addressing the fluctuations for large \texorpdfstring{N}{} simulations}

\begin{figure}[t]
  \centering
  {\includegraphics[width=0.62\linewidth]{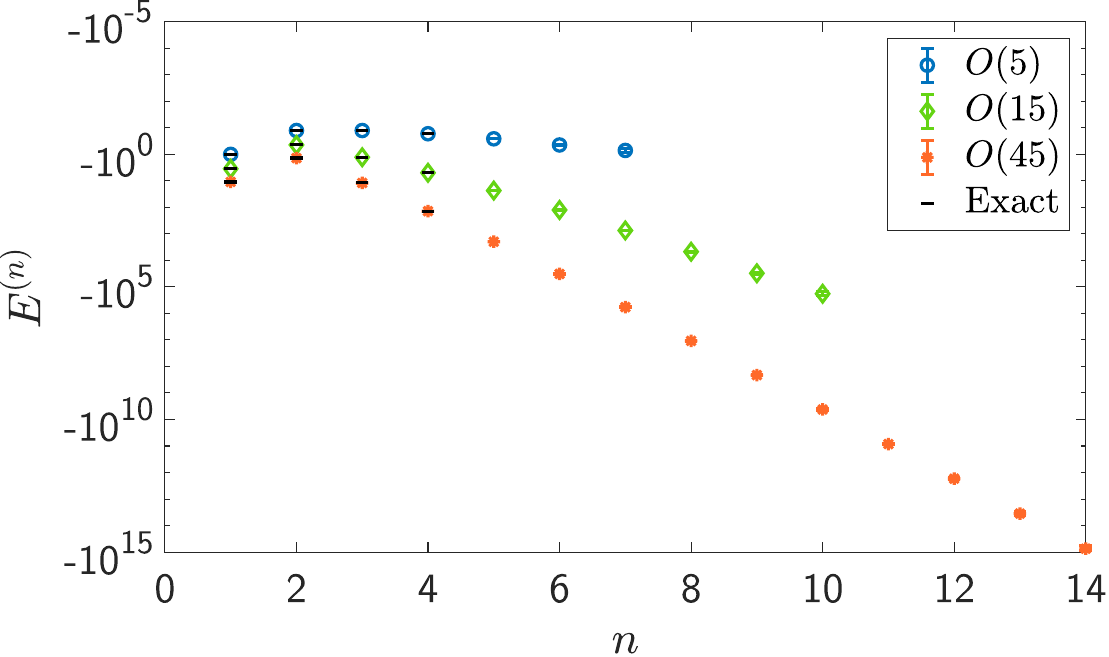}}
  \caption{\textbf{Perturbative corrections to the energy density}, see Eq.~\eqref{eq:energy}. We display with colored symbols the NSPT perturbative computations at different values of $N$ (see the legend), while the analytically known results are reported in black. High-order NSPT simulations are reliable for larger values of $N$, extending loop computations to higher perturbative orders as $N$ increases.}
  \label{fig:comparison}
\end{figure}

We simulate the system in Eq.~\eqref{eq:PF_PT} using the NSPT algorithm for different values of $N$ on $2D$ square lattices of size $V = 20\times 20$. Eqs.~\eqref{eq:orderByOrderLangevin} are numerically integrated using the Euler scheme, considering different values of the step size $\Delta\tau = 0.01, 0.0075, 0.005, 0.0035, 0.0025, 0.0018$. Additionally, systematic effects arising from the numerical integration have been removed by considering first- and second-order contributions in $\Delta\tau$ (see, for example, Eq. (31) of \cite{Baglioni2024c}). Following Eq.~\eqref{eq:PF_PT}, the theory exhibits a zero-mode that requires regularization. A common approach is to introduce an infrared regulator $\lambda$ to be removed at the end of the simulations, in the same spirit of Eq.~\eqref{eq:PF_PT}. However, since no suitable regularizations can be obtained with small values of $\lambda$, we choose to directly remove the zero-mode from the field configuration at each NSPT update. In this way, lattice perturbation theory predictions are recovered in the limit of infinite volume \cite{DiRenzo2004}.

Comparing analytical results and NSPT computations, we found very good agreement even on small lattices up to four loops. In Fig.~\ref{fig:comparison}, we show loop corrections to the energy as a function of the loop order $n$ for the cases $N=5, 15,$ and $45$. In the low perturbative order region, we reported NSPT predictions for all the models considered. On the contrary, for high perturbative orders, progressively larger perturbative corrections can be reliably computed with NSPT only for models with larger $N$, owing to well controlled fluctuations in those cases. As shown in Fig.~\ref{fig:comparison} we computed loop corrections up to the seventh order for the $O(5)$ model, and up to the tenth order for the $O(15)$ model; in addition, we extended the computation up to the fourteenth order for the $O(45)$ model, significantly expanding the analytically known results.

\begin{figure}[t]
  \centering
  {\includegraphics[width=1.\linewidth]{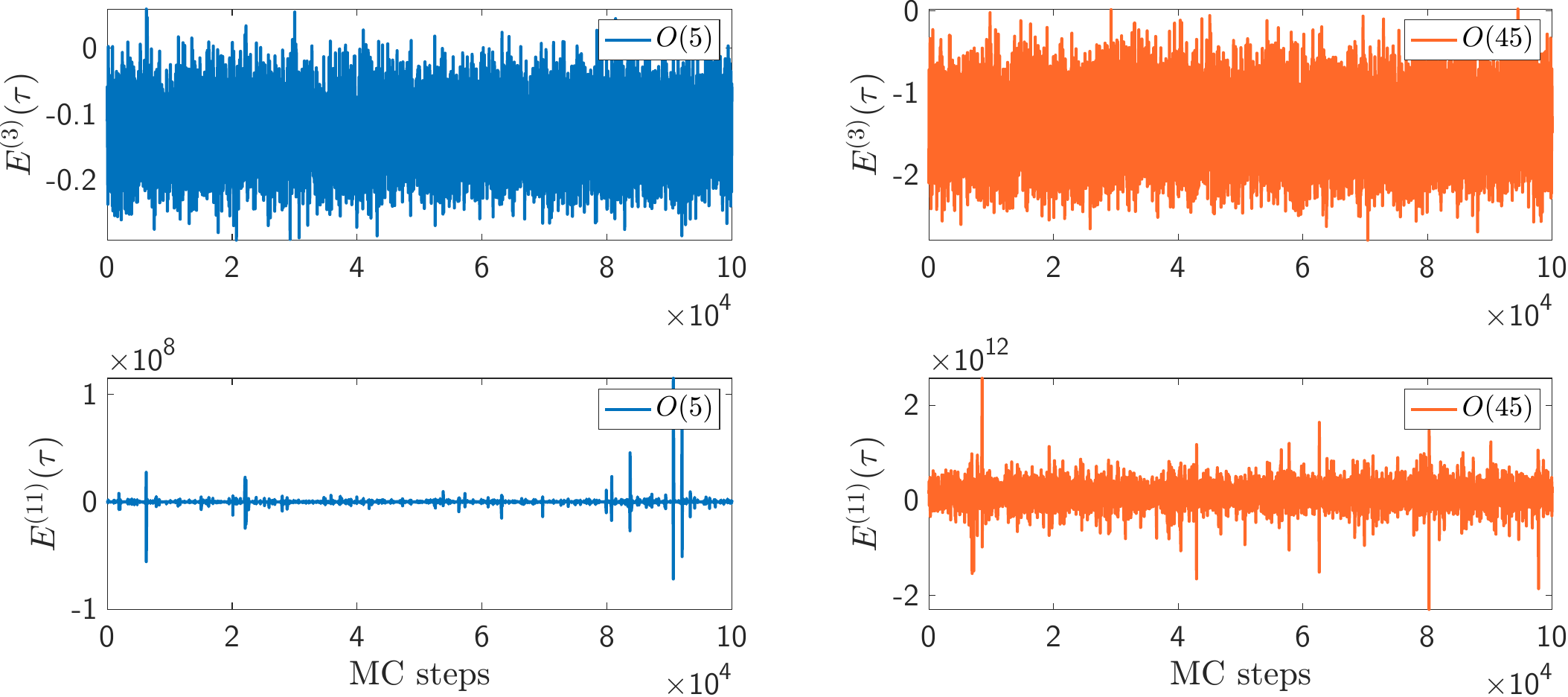}}
  \caption{\textbf{Comparison between small $N$ and large $N$ NSPT evolutions}. Time histories for the $O(5)$ model are reported with blue lines, while evolutions for the $O(45)$ model are shown with orange lines. Both simulations are performed at the same stochastic time step $\Delta\tau = 0.01$.}
  \label{fig:signals}
\end{figure}

Large fluctuations emerge at relatively low perturbative orders in the $O(5)$ model, plaguing the NSPT signal with huge spikes. A paradigmatic case is illustrated in Fig.~\ref{fig:signals}: here we display signals from NSPT simulations at low (high) perturbative orders in the first (second) row, both for the $O(5)$ and $O(45)$ models. While at low perturbative order ($n=3$, first row) the signals appear quite similar to a Gaussian one, this is not the case at perturbative order $n=11$: during the stochastic evolution the $O(5)$ model distributions display extended and not exponentially suppressed tails. In this scenario, NSPT distributions deviate significantly from a Gaussian distribution, and no stable mean and standard deviation can be computed (further details will be provided later). Conversely, in the large $N$ simulations huge spikes are substantially reduced, and the oscillations are entirely manageable at the same time scale (see the bottom-right plot of Fig.~\ref{fig:signals}). It is worth noticing that these findings align with recent NSPT simulations of the twisted Eguchi-Kawai model, where a trend towards Gaussianization of NSPT distributions has been observed at low perturbative orders in the large $N$ limit \cite{GonzlezArroyo2019}.

\subsection{Cumulative mean and standard deviation}

\begin{figure}[t]
  \centering
  {\includegraphics[width=1.\linewidth]{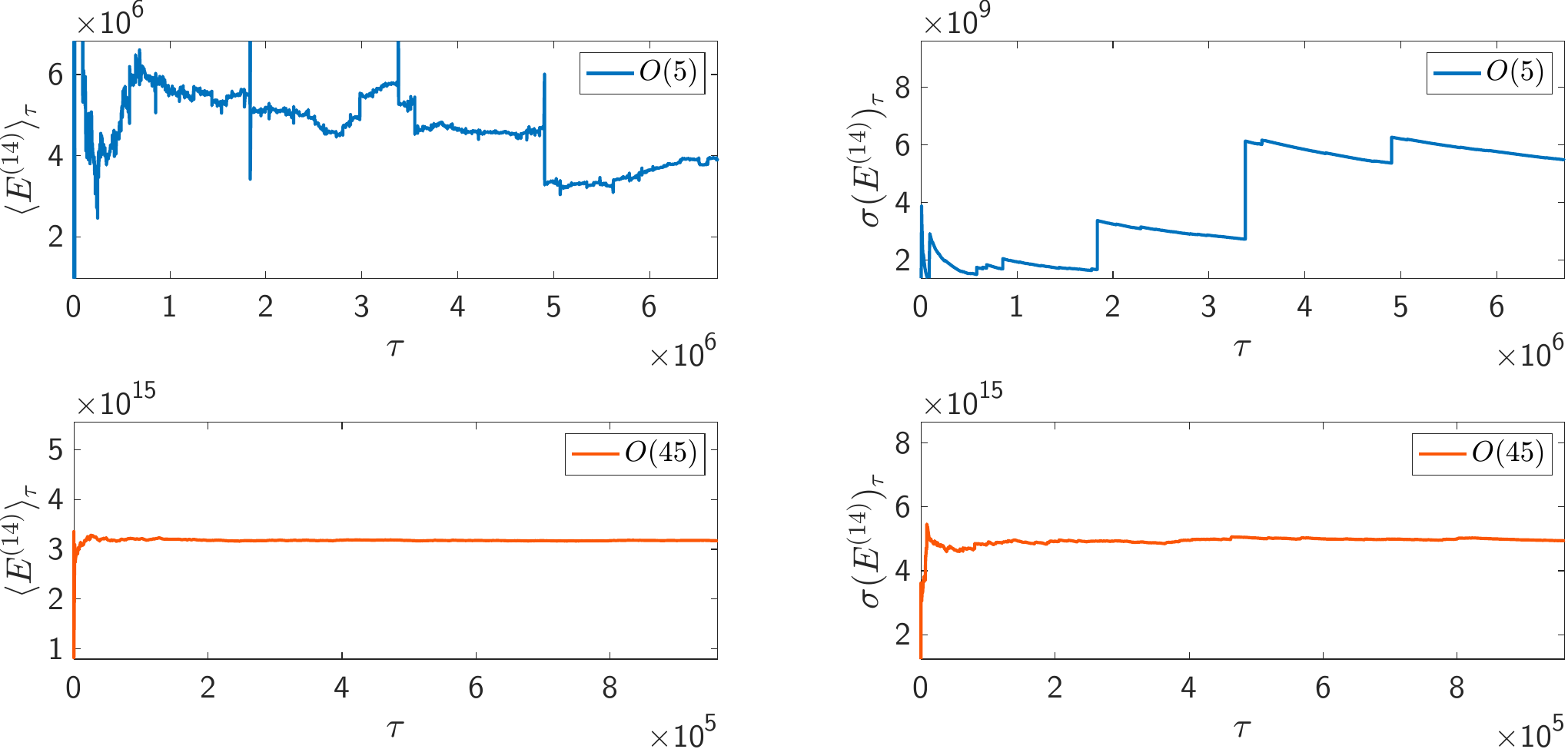}}
  \caption{\textbf{Cumulative mean and standard deviation.} Simulations refer to the same time step $\Delta\tau=0.01$. While different time scales are adjusted to display approximately the same amount of computational time, the $y$-axes are set to show the same relative variation for different $O(N)$ models.}
  \label{fig:cumulative}
\end{figure}

Continuum stochastic time extrapolations require reliably determined order-by-order mean and standard deviation. Since in the context of NSPT simulations, order-by-order distributions are populated with new measurements as the total stochastic time $\tau = \Delta\tau \cdot n_{\text{steps}}$ increases, an effective way to study the onset of fluctuations in the determination of mean and standard deviation is to consider their cumulative definition \cite{Alfieri2000}:
\begin{align}
	\label{eq:cumMean}
	\langle E^{(n)} \rangle_\tau = \frac{1}{\tau}\sum_{i=1}^{\tau}E^{(n)}_i, \quad\quad\quad \text{(cumulative average)}
\end{align}
\begin{align}
	\label{eq:cumStd}
	\sigma(E^{(n)})_\tau = \sqrt{\langle E^{(n)^2} \rangle_\tau - \langle E^{(n)} \rangle_\tau^2 } , \quad\quad\quad \text{(cumulative standard deviation)}
\end{align}
For well-explored distributions, the quantities in Eqs.~\eqref{eq:cumMean}-\eqref{eq:cumStd} are expected to approach an asymptotic constant value at large simulation time $\tau \gg 1$. We note that different $O(N)$ models are related with different computational costs, due to the varying number of degrees of freedom to be updated; therefore, for a fair comparison, we discuss time evolution for Eqs.~\eqref{eq:cumMean}-\eqref{eq:cumStd} considering roughly the same amount of computational time rather than the same statistics. In Fig.~\ref{fig:cumulative} we show cumulative mean and standard deviation at high perturbative order, namely $n=14$, both for the $O(5)$ and $O(45)$ model (the color code is the same as Fig.~\ref{fig:comparison} and Fig.~\ref{fig:signals}). For $N=5$, as a consequence of the onset of fluctuations, the determination of the mean suffers from severe spikes at all simulation time scales (see left plot on the first row). In this case, we have to deal with considerable indeterminacy of the mean, which persists even with a large amount of statistics. This effect is amplified for the standard deviation (see the upper right plot of the same figure), which accounts only for positive contributions (see Eq.~\eqref{eq:cumStd}). In this case, we cannot determine whether the standard deviation of the stochastic process can be determined at all. 
In the second row of Fig.~\ref{fig:cumulative}, we display the same quantities for the $O(45)$ model. At $N=45$, NSPT simulations are stable, providing a safe determination for both the mean and the standard deviation. Additionally, in the large $N$ limit, the standard errors are observed to scale properly as $\sim 1/\sqrt{N_{\text{samples}}}$. This scenario is independent of the choice of the stochastic time step $\Delta\tau$, and consequently, the extrapolation to the continuum stochastic time is well-defined in the large $N$ limit.

\subsection{Relative error scaling}

In this section, we present a preliminary numerical study of the scaling of relative errors. Specifically, we investigate the relative error as a function of the parameter $N$, defined as
\begin{equation}
	\label{eq:relativeError}
	\bar{\Delta}^{(n)}_N = \bigg|\frac{\delta E^{(n)}}{E^{(n)}}\bigg|_{N} \cdot \Gamma(N) \, .
\end{equation}
In the above equation, we are considering the ratio of the $n$-th statistical error $\delta E^{(n)}$ to its associated perturbative correction $E^{(n)}$ at fixed values of $N$. For each $O(N)$ model, we introduce a corrective prefactor $\Gamma(N)$. In Eq.~\eqref{eq:relativeError}, $\Gamma(N)$ adjusts the relative error based on the corresponding statistics, so that Eq.~\eqref{eq:relativeError} broadly represents the relative error per unit of statistics, since the \textit{bare} relative errors are influenced by the differing amounts of statistics for different $N$ values. In other words, we are left only with genuine $N$ effects. It is important to highlight that relative errors, as defined in Eq.~\eqref{eq:relativeError}, are affected by all the effects stemming from auto and cross-correlations, which are addressed consistently during the continuum stochastic time extrapolation (we refer the reader to the appendix of \cite{Baglioni2024c} for further details). 

On very general grounds, the relative errors in Eq.~\eqref{eq:relativeError} are expected to exhibit a monotonically decreasing trend with increasing $N$. Nevertheless, the ratio computed in Eq.~\eqref{eq:relativeError} is based on our \textit{best} estimations of $E^{(n)}$ and $\delta E^{(n)}$: given the issues at high orders, some of these estimations may prove to be unreliable.

\begin{figure}[t]
  \centering
  {\includegraphics[width=1.\linewidth]{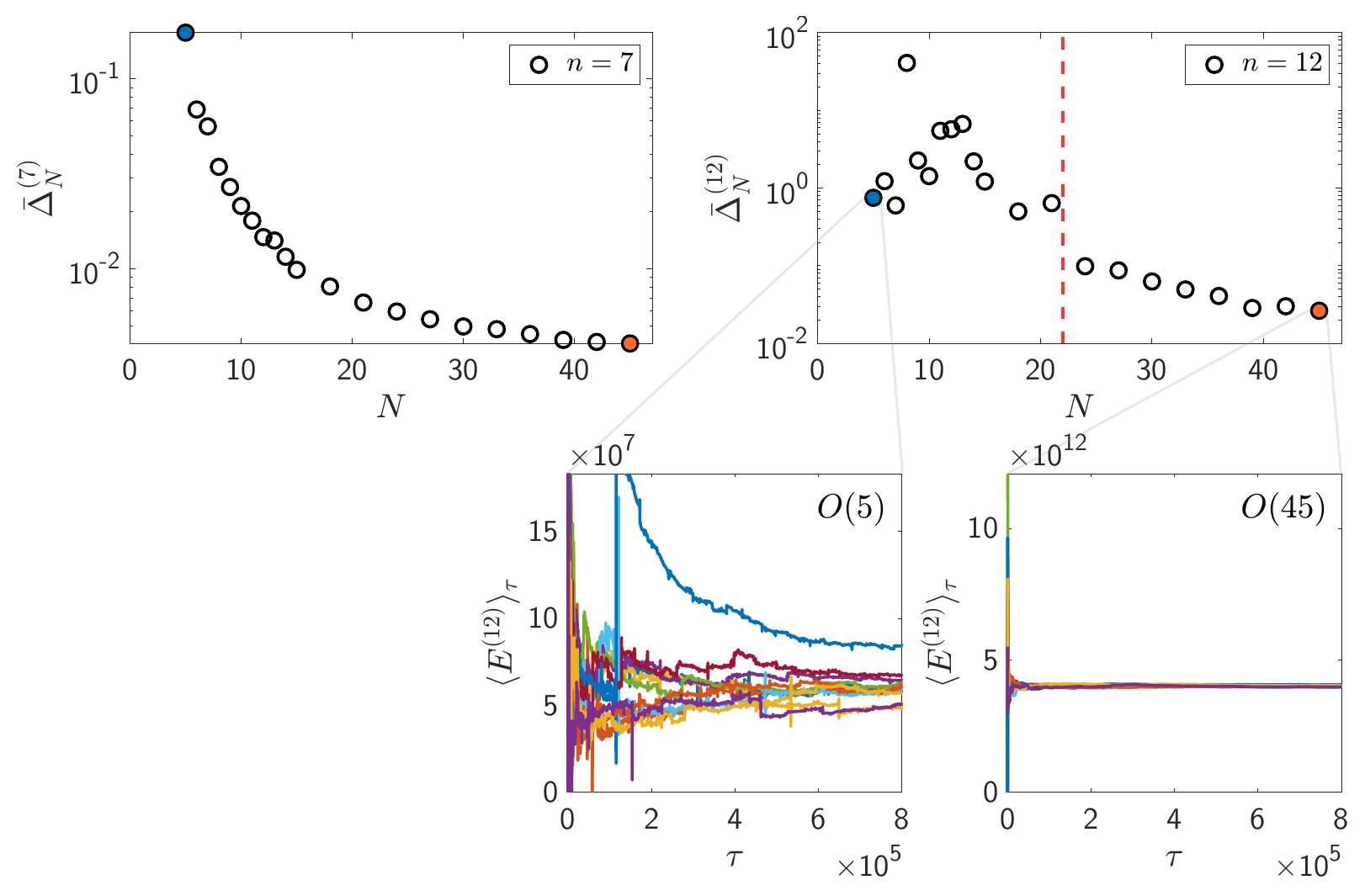}}
  \caption{\textbf{Scaling of relative errors with respect to $N$}: In the first row we depict the relative errors scaling as a function of $N$ (see the definition in Eq.~\eqref{eq:relativeError}), for perturbative orders $n=7$ and $n=12$ (respectively in the upper left and upper right plot). Colored markers represent relative errors for $N=5$ and $N=45$ (with the same color code as before). The dashed red line (right plot) delineates the region where the expected scaling is observed. In the second row, we display various determinations of the cumulative mean from different independent runs, both for the $O(5)$ and $O(45)$ models.}
  \label{fig:relativeError}
\end{figure}

In Fig.~\ref{fig:relativeError} - first row, we show relative errors for different values of $N$. Filled colored markers represent the $O(5)$ and $O(45)$ models (with the same color code as before). Our hypothesis is well verified in the low perturbative order region (see the upper left plot). For $n=7$, no deviations from the expected trend are observed for all the $O(N)$ models. However, it is not surprising that at high perturbative orders, hypothesis violations emerge in the small $N$ region. For example, at perturbative order $n=12$ (see the upper right plot), no monotonically decreasing trend can be established for $N\lesssim22$ (notice the red dashed line). For $N\gtrsim 22$, the expected trend is recovered, and we have no evidence of the effects of fluctuations in such a region. It is worth noticing that our simulations clearly indicate that for increasing perturbative orders, we can always find a large enough $N$ where the order-by-order distributions are easy to explore without reaching prohibitive simulation times (for a more detailed study, see \cite{Baglioni2024c}). In Fig.~\ref{fig:relativeError} - second row, we depicted the evolution of the cumulative mean at high perturbative order for multiple independent runs: while in the small $N$ region different runs are characterized by very different determinations of the mean (the same holds for the cumulative standard deviation, see \cite{Baglioni2024c}), at large $N$ we have no evidence of such discrepancies. 

Overall, our analysis suggests that NSPT simulations are reliable in the large $N$ limit: as $N$ increases, more perturbative corrections can be safely computed. 

\section{Outlook and perspective}

In this work, we numerically show that NSPT computations are safe in the large $N$ limit of $O(N)$ models: the fluctuations that show up for small $N$ are not there. Notably, this does not mean one has to confront excessively large values of $N$, which would incur prohibitively high computational costs. Specifically, high-order predictions can be obtained for $N\gtrsim 45$, opening up new possibilities for NSPT applications in low-dimensional models. Currently, large $N$ simulations are underway to explore infrared renormalons in $O(N)$ models, with promising preliminary results \cite{Baglioni2024d}. In addition, $O(N)$ are a first step to later tackle $CP(N-1)$ models, which feature non-trivial vacua \cite{zakrzewski1989} suitable for perturbative expansions.

\section *{Acknowledgments}
We thank Petros Dimopoulos for discussions. This work was supported by the European Union Horizon 2020 research and innovation programme under the Marie Sk\l odowska-Curie grant agreement No 813942 (EuroPLEx) and by the INFN under the research project (\textit{iniziativa specifica}) QCDLAT. This research benefits from the HPC (High Performance Computing) facility of the University of Parma, Italy. We acknowledge also the CINECA computing centre, for the availability of high-performance computing resources and support.

\bibliographystyle{JHEP}
\bibliography{reference}

\end{document}